\documentclass[12pt]{article}
\newcommand{\beq}{\begin{eqnarray}}% can be used as {equation} or {eqnarray}
\newcommand{\eeq}{\end{eqnarray}}
\newcommand{\nn}{\nonumber}
\def\){\right)}
\def\({\left( }
\def\]{\right] }
\def\[{\left[ }

\title{A New  Angle on Intersecting Branes in Infinite
Extra Dimensions}

\author{
  Ann E. Nelson$^c$\ \thanks{\tt  anelson@fermi.phys.washington.edu}\\ \\
           Box 351560, Dept. of Physics, University of Washington,\\
 Seattle, WA 98195-1560, USA\\ \\
  }

\begin{document} 
\setlength{\baselineskip}{24pt}
\begin{titlepage}

\maketitle
\begin{picture}(0,0)(0,0)
\put(295,350){UW/PT-99/16}
\end{picture}
\vspace{-36pt}

\begin{abstract}
I construct  solutions to Einstein's equations in 6 dimensions with  
bulk cosmological
constant and intersecting 4-branes. Solutions exist for a continuous
range of 4-brane
tension, 
with long distance gravity  localized to a 3+1 dimensional Minkowski
intersection, provided that the additional tension of the intersection
satisfies
one condition.

\end{abstract}
\thispagestyle{empty}
\setcounter{page}{0}
\end{titlepage}

\section{Introduction}

String/M-theory requires the existence of nine or ten spatial
dimensions. Traditionally it has been assumed that all but three
spatial dimensions are compact, with size comparable to the Planck
length.  Recently it has been proposed that at least some of 
the extra dimensions are compact but very large, 
 with size between a Fermi and a millimeter\footnote{For earlier work
on TeV$^{-1}$ sized new dimensions, see ref. \cite{Antoniadis:1990ew}.},
 but that the fields of
the Standard Model are confined to a $3+ 1$ dimensional subspace, or 
``3-brane''
\cite{Arkani-Hamed:1998rs,Antoniadis:1998ig}. Large new
dimensions in which the standard model fields do not propagate may 
allow for the unification of 
gravity with the other forces and strong quantum gravity effects below
the Planck 
scale \cite{Horava:1996qa,Witten:1996mz,Horava:1996ma},
perhaps as low as a few 
TeV \cite{Arkani-Hamed:1998rs,Antoniadis:1998ig,Lykken:1996fj}.
An alternative picture due to Randall and Sundrum
(RS)\cite{Randall:1999ee} 
also assumes that
the Standard Model is confined to a 3-brane, but allows for a
non-factorizable metric with  $4$ dimensional Minkowski space
embedded in a slice of $5$ dimensional Anti-deSitter (AdS)
space\footnote{These solutions parallel earlier solutions found in 4
dimensions \cite{Cvetic:1992bf} reviewed in ref. \cite{Cvetic:1996vr}.}. The
exponential dependence of the coefficient of the 4 dimensional
metric tensor on an additional
coordinate (the ``warp factor'') can provide a explanation of the 
weak/Planck hierarchy
without a large extra 
dimension \cite{Randall:1999ee,Arkani-Hamed:1999hk,Lykken:1999nb}. Randall and
Sundrum \cite{Randall:1999vf} have pointed out that  such a warp
factor also allows for a non-compact extra dimension, 
with a  normalizable 0-mode of the graviton providing effective 3+1
dimensional Einsteinian gravity at long distances, even though there is a
continuous, gapless spectrum of Kaluza-Klein gravitational
modes\footnote{See 
also  refs.~\cite{Rubakov:1983bz,Gell-Mann:1985if,Randjbar-Daemi:1986wg,Brandhuber:1999hb,Gogberashvili:1999ad}.}.
Such a normalizable zero mode is a general feature of any space-time
metric that has a 3+1 dimensional Minkowski subspace and a warp factor
which 
decreases sufficiently fast in all new directions~\cite{ckn}.
Arkani-Hamed, Dimopoulos, Dvali and Kaloper (ADDK) have extended the RS
work  to  $n$ intersecting $(2+n)$ branes  separating sections of 
$4+n$ dimensional AdS space\footnote{As this work was being
completed a paper appeared which further generalizes the RS solution\cite{Csaki}.}.
Both the RS and the  ADDK solutions require a specific relationship
between the brane tensions and the bulk cosmological constants. In
addition, the ADDK solution requires that there is no additional
contribution to the tension of the intersection from brane-brane
interactions.
In this note I further generalize the ADDK solution for the case of 2
new dimensions, to the case of AdS space containing  $n$ AdS branes
with 
almost arbitrary brane
tension meeting at a 3+1 dimensional Minkowski junction with localized
gravity. One fine-tuning is required as there is a  constraint on the
value of the brane-brane interaction contribution to the tension of
the 
intersection.
\section{AdS Branes in AdS Space Intersecting at Angles}
The following metric is a solution to Einsteins equations in 5+1
dimensions
with a
 bulk cosmological constant and a   
 4-brane located along the $y$ axis.  
\beq
\label{metric}
ds^2&=&\Omega^2(\eta_{\mu\nu}dx^\mu dx^\nu+dy^2+dz^2)\\
\Omega&=&1/\[1+k\(y\cos{\phi}+|z|\sin{\phi} \)\]   
 \nn
\eeq
Here the usual 3+1 dimensions are denoted by $x^\mu$ and the 2 new
dimensions by $y$ and $z$. 
The brane glues together 2 semi-infinite 
slices of Anti-deSitter (AdS) space. In
units of the 6 dimensional Newton's constant, the bulk
cosmological constant is $10 k^2$ while the brane tension is 
$8 k\sin\phi$. For
$\phi=90^o$ 
this solution trivially generalizes the Randall-Sundrum metric to one
additional dimension, and the induced metric on the brane is
5 dimensional Minkowski. Otherwise the induced metric is 5 dimensional AdS.
Note that the angle $\phi$ is determined by the
relationship between 2 apparently nondynamical parameters--the brane 
tension and the bulk cosmological
constant. 
Now consider $n$   branes, intersecting on a  3+ 1 dimensional
subspace.
Between adjacent  pairs of branes is a wedge of 6 dimensional AdS space.
The metric within one wedge may be written as in eq.~\ref{metric}
in a coordinate system where there is one semi-infinite brane located along the
positive $y$ axis, and another extending from the origin in the
$z>0$ half-plane along
the line 
$y\cos(2\phi) -z\sin(2\phi)=0$. The metric is reflection symmetric
about the line $y\cos(\phi) -z\sin(\phi)=0$.
 For 
$\phi=45^o$, 
a  solution pasting together 4 such wedges was found by ADDK~\cite{Arkani-Hamed:1999hk}.  The
ADDK solutions require a fine-tuned relation between the bulk
cosmological constant and the brane tension, 
and also that there be no
brane-brane interaction tension localized at the intersection. 
Generally, however, one would not expect $\phi=180^o/n$ for
integer $n$. However
one could still patch together $n$   sections of six
dimensional
AdS space along $n$  AdS 4-branes,
by allowing  a global ``deficit angle''  (in analogy with the case of the gauge
string metric \cite{Vilenkin:1981zs,Gott:1985ef}) of $360^o-2n\phi$. 
In a spacetime patch around  each brane, in a coordinate system rotated such that the brane is located
at along the positive $y$ axis, the metric is given by
eq.~\ref{metric}.
Thus this patched together metric is automatically a solution to Einstein's equations
everywhere except in the core of the intersection. For a global
solution to exist  
 the brane-brane interaction must produce a specific
  contribution to the tension of the
intersection region, needed to match the global deficit angle.
Thus it still appears that one fine tuning of a nondynamical parameter
(or a remnant
supersymmetry)
is required. Perhaps 
a better understanding of brane interactions will shed some light on the
cosmological constant question.

\section{Outlook on Extra Dimensions and the Cosmsological Constant}

The idea that noncompact higher dimensions with nonfactorizable
spacetime
might be consistent with 
our successful effective long distance theories and  explain why
the cosmological constant is zero has been around for a long time
\cite{Rubakov:1983bb,Rubakov:1983bz}. The hope is that the metric can
adjust such that in equilibrium an apparent cosmological term will
only 
affect the
part of the metric corresponding to the additional dimensions. For
instance in the case of a gauge string, the vacuum energy of any
effective 
degrees of freedom which
live only on the string contribute to the string tension, but the
induced metric on the string remains Minkowski, with any change in the
string
tension simply changing the deficit angle. Extra
dimensional theories typically have new light degrees of freedom. New
light fields
are needed to understand any
cosmological constant 
adjustment mechanism from an effective field theory point of view. 
It is conceivable that new degrees of freedom associated with an
extra-dimensional metric somehow cleverly avoid the usual 
problems of adjustment mechanisms \cite{Weinberg:1989cp}. Since
the effective four dimensional Newton's constant typically depends on the extra
dimensional metric, perhaps a higher dimensional
configuration can be found where apparent low energy contributions to
the four dimensional cosmological constant (e.g. the QCD phase
transition)  actually just renormalize Newton's constant.
The configuration described in this note does not, however, seem so
clever.

\noindent\medskip\centerline{\bf Acknowledgments}
While this work was in progress ref.~\cite{Csaki} appeared, with
overlapping 
results and speculations. 
I thank David Kaplan,  Patrick Fox, and Andrew Cohen for useful
conversations
and the Aspen Center for Physics for hospitality during the
inception of this work.
The work was supported in part by DOE grant DE-FG03-96ER40956.

\bibliography{brane}
\bibliographystyle{utcaps}

\end{document}